\documentstyle[12pt,epsfig,a4]{article} 
\newcommand{\vs}{\vspace{-0.25cm}}
\begin{document} 

\begin{center}
{\Large{\bf In-medium chiral condensate \\ beyond linear density approximation}
}\footnote{Work supported in part by BMBF, GSI and by the DFG cluster of
excellence: Origin and Structure of the Universe.}

\medskip

N. Kaiser, P. de Homont, and W. Weise\\

\smallskip

{\small Physik Department, Technische Universit\"{a}t M\"{u}nchen,
    D-85747 Garching, Germany}
\end{center}

\medskip

\begin{abstract}
In-medium chiral perturbation theory is used to calculate the density 
dependence of the quark condensate $\langle\bar qq\rangle$. The corrections 
beyond the linear density approximation are obtained by differentiating the 
interaction contributions to the energy per particle of isospin-symmetric 
nuclear matter with respect to the pion mass. Our calculation treats 
systematically the effects from one-pion exchange (with $m_\pi$-dependent 
vertex corrections), iterated $1\pi$-exchange, and irreducible $2\pi$-exchange 
including intermediate $\Delta(1232)$-isobar excitations, with Pauli-blocking 
corrections up to three-loop order. We find a strong and non-linear dependence 
of the ``dropping'' in-medium condensate on the actual value of the pion (or 
light quark) mass. In the chiral limit, $m_\pi=0$, chiral restoration appears 
to be reached already at about $1.5$ times normal nuclear matter density. By 
contrast, for the physical pion mass, $m_\pi = 135\,$MeV, the in-medium 
condensate stabilizes at about $60\%$ of its vacuum value above that same 
density. Effects from $2\pi$-exchange with virtual $\Delta(1232)$-isobar 
excitations turn out to be crucial in generating such pronounced deviations 
from the linear density approximation above $\rho_0$. The hindered tendency 
towards chiral symmetry restoration provides a justification for using pions 
and nucleons as effective low-energy degrees of freedom at least up to twice
nuclear matter density.
\end{abstract}
\medskip 

PACS: 12.38.Bx, 21.65.+f\\
Keywords: In-medium chiral condensate, long-range correlations in nuclear
matter from one- and two-pion exchange. 
\medskip
\section{Introduction and framework} 
The quark condensate $|\langle 0|\bar q q|0\rangle|$ is an order parameter of 
spontaneous chiral symmetry breaking in QCD. With increasing temperature the 
quark condensate decreases (or ''melts''). For low temperatures this effect 
can be systematically calculated in chiral perturbation theory. At three-loop 
order \cite{gerber} the estimate $T_c \simeq 190\,$MeV for the critical 
temperature, where chiral symmetry will eventually be restored, has been 
found. This extrapolated value of $T_c$ is remarkably consistent with $T_c = 
(192\pm 8)\,$MeV \cite{cheng} obtained in numerical simulations of full QCD on 
the lattice, although this result is still under debate. It has subsequently 
been criticized \cite{aoki} with respect to the reliability of the continuum
extrapolation performed in ref.\cite{cheng}. In fact the QCD  chiral ''phase 
transition'' is merely a non-singular cross-over, with the transition 
temperature (deduced from the peak of the chiral susceptibility) lying in the 
broad range $T_c = (160\pm 29)\, $MeV according to ref.\cite{aoki}. 

The chiral condensate  $|\langle\bar q q\rangle|$ drops also with increasing 
baryon density. Presently, it is not feasible to study this phenomenon
rigorously in lattice simulations of QCD due to the problems arising from the 
complex-valued Euclidian Fermion determinant at non-zero baryon chemical 
potential. As an alternative, the density dependence of the quark condensate  
$\langle\bar q q\rangle(\rho)$ can be extracted by exploiting the
Feynman-Hellmann theorem applied to the chiral symmetry breaking quark mass
term $m_q\, \bar qq$ in the QCD Hamiltonian. The leading linear term in the
nucleon density $\rho$ is then readily derived by differentiating the energy
density of a nucleonic Fermi gas, $\rho M_N + {\cal O}(\rho^{5/3})$, with
respect to the light quark mass $m_q$. This introduces the nucleon sigma-term 
$\sigma_N = \langle N|m_q\,\bar q q|N\rangle = m_q\, \partial M_N /\partial
m_q =(45\pm 8) \,$MeV \cite{sigma} as  the driving term for the density
evolution of the chiral condensate. Following this simple linear density
approximation one would naively estimate that chiral  symmetry gets restored at
$(2.5-3)\rho_0$, with $\rho_0 = 0.16\,$fm$^{-3}$ the nuclear matter saturation
density.  

Corrections beyond the linear density approximation arise from the 
nucleon-nucleon interactions which transform the nucleonic Fermi gas into a 
nuclear Fermi liquid. These corrections have been studied in one-boson 
exchange models of the NN-interaction combined with the relativistic 
Dirac-Brueckner approach to nuclear matter \cite{cohen,ko}. Knowledge of the 
quark mass derivatives of the various meson masses and coupling constants is  
required in order to quantify the interaction effects on the in-medium 
condensate. For the vector and scalar bosons it has been assumed that their 
sigma-terms scale linearly with that of the nucleon, i.e. $m_q\,\partial m_{V,S} 
/\partial m_q= \sigma_N m_{V,S}/M_N$. In the further development it has been 
demonstrated in ref.\cite{rolf} that the higher order corrections (in density) 
depend sensitively on the interpretation  of the isoscalar scalar ''$\sigma
$''-boson (which is responsible for the central NN-attraction in one-boson 
exchange models) and its substructure. Within modest variations of an unknown 
parameter $0\leq C_S\leq 1$ both an accelerated and a hindered tendency
towards chiral restoration are possible. The way out of this dilemma is to 
replace the fictitious ''$\sigma$''-boson exchange by realistic two-pion 
exchange processes. 

Because of the Goldstone boson nature of the pion with its characteristic mass
relation, $m_\pi^2 \sim m_q$, the pion-exchange dynamics in nuclear matter 
plays a particularly important role for the in-medium quark condensate. A
first step in this direction was made in ref.\cite{lutz} where realistic 
saturation of nuclear matter could be obtained from the iteration of 
$1\pi$-exchange plus a short-range NN-contact interaction to second order. It 
was found that the pionic interaction effects (with well-known quark mass 
derivative) counteract the reduction of the condensate from the leading linear 
term in density. Actually, if one restricts oneself to the density region 
$\rho \leq \rho_0 = 0.16\,$fm$^{-3}$, then all existing calculations agree
that the deviations from the linear density approximation are relatively small 
and practically get masked by the uncertainty of the empirical nucleon 
sigma-term, $\sigma_N = (45\pm 8)\,$MeV.   

The chiral approach to nuclear matter has been extended and improved in
refs.\cite{nucmat,deltamat} by systematically including effects from 
irreducible $2\pi$-exchange together with excitations of virtual $\Delta(1232)
$-isobars. The physical motivation for such an extension is threefold. First, 
the spin-isospin-$3/2$ $\Delta(1232)$-resonance is the most prominent feature 
of low-energy $\pi N$-scattering. Secondly, it is well known that the $2\pi
$-exchange between nucleons with excitations of virtual $\Delta$-isobars 
generates the medium- and long-range components of the isoscalar central 
NN-attraction \cite{2pidel}, which is simulated by the scalar ''$\sigma
$''-boson in the one-boson exchange models. Thirdly, the 
delta-nucleon mass-splitting $\Delta = 293\,$MeV is of the same size as the 
Fermi momentum $k_{f0} =263\,{\rm MeV} \simeq 2m_\pi$ at nuclear matter
saturation density. Therefore pions and $\Delta$-isobars should both be 
treated as explicit degrees of freedom in the nuclear many-body problem. It 
has been found in ref.\cite{deltamat} that the inclusion of the chiral $\pi 
N\Delta$ dynamics significantly improves e.g. the  momentum-dependence of the
(real) single-particle potential $U(p,k_f)$ and the isospin properties (as
revealed by the density-dependent asymmetry energy  $A(k_f)$ and the neutron
matter equation of state). Moreover, it guarantees spin-stability of nuclear 
matter \cite{spinstab}.  

The purpose of the present paper is to investigate the in-medium chiral
condensate (beyond the linear density approximation) in this extended and
improved framework for nuclear matter where interaction effects are (almost)
exclusively given by one- and two-pion exchange according to the rules of 
chiral symmetry. Since the pion mass $m_\pi$ appears as an explicit variable
in our calculation we can also study how the  ''dropping'' in-medium
condensate evolves from the chiral limit, $m_\pi = 0$, to the real world with
its fixed amount of explicit chiral symmetry breaking, $m_\pi = 135\,$MeV. We 
find that the in-medium condensate behaves very differently in both cases, 
with drastic consequences for nuclear physics in the chiral limit. 

Our starting point is the Feynman-Hellmann theorem which relates the in-medium
quark condensate  $\langle \bar q q\rangle(\rho)$ to the quark mass derivative 
of the energy density of isospin-symmetric spin-saturated nuclear matter. 
Using the Gell-Mann-Oakes-Renner relation $m_\pi^2 f_\pi^2 = -m_q  \langle 0|
\bar q q|0 \rangle$ one finds for the ratio of the in-medium to the vacuum 
quark condensate: 
\begin{equation} { \langle \bar q q\rangle(\rho)\over \langle 0|\bar q q|0 
\rangle} = 1 -{\rho \over f_\pi^2} \bigg\{ {\sigma_N \over m_\pi^2} \bigg(1 
-{3k_f^2 \over 10 M_N^2} +{9k_f^4 \over 56 M_N^4}   \bigg) + D(k_f) \bigg\}\,,
\end{equation}
with the Fermi momentum $k_f$ related to the nucleon density $\rho = 
2k_f^3/3\pi^2$ in the usual way. The term proportional to $\sigma_N=\langle N|
m_q\,\bar q q|N\rangle =m_q \,\partial M_N/\partial m_q$ comes from the 
noninteracting Fermi gas including kinetic energy contributions expanded up to
order $M_N^{-3}$. Interaction contributions beyond the linear density
approximation are collected in the function: 
\begin{equation}D(k_f) =  {1\over 2m_\pi }  {\partial \bar E(k_f) \over
 \partial  m_\pi} \,, \end{equation}
defined as the derivative of the interaction energy per particle $\bar E(k_f)$
with respect to $m_\pi^2$. We mention here that $f_\pi$ denotes the pion decay 
constant in the chiral limit and $m_\pi^2$ stands for the leading linear term 
in the quark mass expansion of the squared pion mass. With this convention the 
Gell-Mann-Oakes-Renner relation, $m_\pi^2 f_\pi^2 = -m_q \langle 0|\bar q q|0 
\rangle$, becomes exact and $\langle 0|\bar q q|0 \rangle$ is the vacuum 
condensate in the chiral limit.   

\section{Pion mass derivative of the interaction energy} 
In this section we present analytical expressions for the contributions to
the derivative function $D(k_f)$ as given by various one- and two-pion 
exchange diagrams. Taking the $m_\pi^2$-derivative is a straightforward
procedure since we can borrow here heavily from the explicit expressions for 
the diagrammatic contributions to $\bar E(k_f)$ written down in our previous 
works \cite{nucmat,deltamat}.
\subsection{One-pion exchange Fock term}
We are working to three-loop order in the energy density. At that order one
encounters also pion-loop corrections to the pion-nucleon vertex. As a 
consequence, the $1\pi$-exchange Fock term, eq.(6) in ref.\cite{nucmat}, needs 
to be multiplied by the following renormalization factor: 
\begin{eqnarray} \Gamma(m_\pi) &=& 1+{g_A^2 m_\pi^2 \over (2\pi f_\pi)^2}\bigg[
 4 \gamma + 1-2\ln{m_\pi \over \lambda} \bigg] + {g_A^2 \over 3\pi^2 f_\pi^2} 
\Bigg\{ {\pi m_\pi^3\over \Delta} - {m_\pi^2 \over 2}\nonumber \\ &&+(3m_\pi^2 
-2\Delta^2)\ln{m_\pi \over 2 \Delta } -{2\over \Delta} (\Delta^2-m_\pi^2)^{3/2}
\ln {\Delta +\sqrt{\Delta^2-m_\pi^2} \over m_\pi} \Bigg\} \nonumber \\ && + 
{9g_A^2 \over (4\pi f_\pi)^2} \Bigg\{m_\pi^2 + (4\Delta^2-2m_\pi^2)
\ln{m_\pi \over 2  \Delta } + 4 \Delta \sqrt{\Delta^2-m_\pi^2} \ln {\Delta 
+\sqrt{\Delta^2-m_\pi^2} \over m_\pi} \Bigg\} \,,
\end{eqnarray}
with $g_A$ the nucleon axial vector coupling constant in the chiral limit. The
last two terms, depending on the delta-nucleon mass splitting $\Delta = 293\, 
$MeV, arise from pion-loop diagrams involving the $\Delta(1232)$-isobar. The 
low-energy constant $\gamma(\lambda)$ takes care of the (empirical) 
Goldberger-Treiman discrepancy. It is determined for any choice of the 
regularization scale $\lambda$ by the condition $(g_{\pi N}/M_N)_{\rm  phys}=
\sqrt{\Gamma(m_\pi)} \, g_A/f_\pi$,  where ''phys'' denotes physical values. 

The $1\pi$-exchange Fock diagram (with unrenormalized coupling constant) 
including the relativistic $1/M_N^2$-correction leads to the expression:   
\begin{eqnarray} 
D(k_f)^{(1\pi)} &=& {9 g_A^2 m_\pi \over (8\pi f_\pi)^2} \bigg\{  {1\over 2u} 
-u+ 2 \arctan 2u -{1+8u^2 \over 8 u^3} \ln(1+4u^2) +{m_\pi^2 \over 15M_N^2}  
\nonumber \\ && \times \bigg[ {1\over u}+ {21u \over 2}+ {20 u^3 \over 3} -
\bigg( {25\over 4}+9u^2 \bigg)  \arctan 2u -{1 \over 4 u^3} \ln(1+4u^2) \bigg]
\bigg\} \,,  \end{eqnarray}
where we have introduced the useful dimensionless variable $u= k_f/m_\pi$. For
reasons of consistency with the loop expansion the renormalization factor 
$\Gamma(m_\pi)=1 +{\cal O}(m_\pi^2)$ must be applied only to the static term 
in eq.(4), and the $m_\pi^2$-derivative of $\Gamma(m_\pi)$ must be multiplied  
only with the static $1\pi$-exchange energy per particle. The necessity for 
this procedure will become clear in subsection 2.6 when discussing the chiral 
limit $m_\pi \to 0$.  

\begin{figure}
\begin{center}
\includegraphics[scale=0.9,clip]{selffig.epsi}
\end{center}
\vspace{-0.2cm}
{\it Fig.\,1: One-pion exchange Fock diagram with pion selfenergy. Its 
combinatoric factor is 1/4.}
\end{figure}

The loop diagram with a tadpole on the exchanged pion (see Fig.\,1) generates 
a momentum-independent pion selfenergy (i.e. a pion mass shift). Its 
contribution to the function $D(k_f)$ reads: 
\begin{eqnarray} D(k_f)^{(\pi-\rm self)} &=& {9 g_A^2 m_\pi^3 \over 4(4\pi
f_\pi)^4} \Bigg\{ \bigg(32\pi^2 l_3^r +\ln{m_\pi \over \lambda} \bigg) \bigg[
{2\over u} -2u + 5\arctan 2u -{1+6u^2 \over 2u^3} \ln(1+4u^2)\bigg]  \nonumber 
\\ && +{1 \over 4u} -{u \over 2} + \arctan 2u - {1+8u^2 \over 16u^3}
\ln(1+4u^2) \Bigg\} \,, \end{eqnarray}
with the low-energy constant $l_3^r(\lambda)$ determined by the relation $\bar 
l_3 = -64\pi^2 l_3^r(\lambda) -2\ln(m_\pi /\lambda)\simeq 3$ \cite{cola}. 
Although this contribution is very small it has to be kept for reasons of 
consistency.

\subsection{Iterated one-pion exchange}
The Hartree diagram from iterated $1\pi$-exchange with two medium insertions
(see Fig.\,3 and eq.(7) in ref.\cite{nucmat}) leads to the expression:
\begin{equation} 
D(k_f)^{({\rm it},H2)} ={3\pi g_A^4M_N m^2_\pi \over 5(4\pi f_\pi)^4} \bigg\{  
{63\over 8u} -{193u \over 4}+ (60+16u^2) \arctan 2u -{7\over
  32 u^3}(9+100u^2) \ln(1+4u^2)\bigg\} \,, \end{equation}
and the corresponding Fock diagram with two medium insertions gives:
\begin{eqnarray}
D(k_f)^{({\rm it},F2)} &=&{3\pi g_A^4 M_N m_\pi^2\over (4\pi f_\pi)^4 u^3} 
\int_0^u \!\!dx \,{x(u-x)^2(2u+x)\over (1+2x^2)^2} \bigg[(2+8x^2+16
x^4)\nonumber \\ &&\times(\arctan x-\arctan 2x)+12x^3+4x+{x\over 1+x^2}\bigg]
\,. \end{eqnarray}
In our way of organizing the many-body calculation, the Pauli blocking
corrections are represented by diagrams with three medium insertions. The
contribution of the Hartree diagram with three medium insertions to the
function $D(k_f)$ can be reduced to a one-parameter integral:
\begin{eqnarray}
D(k_f)^{({\rm it},H3)}&=& {9g_A^4 M_N m_\pi^2\over (4\pi f_\pi)^4 u^3}\int_0^u 
\!\! dx\,\bigg[2u x +(u^2-x^2)\ln{u+ x \over  u-x}\bigg] \bigg\{4x(x-u)-{x^2
\over 1+4x^2} \nonumber \\ && +{u(x+u)\over 2[1+(u+x)^2]} + {u(x-u)\over 2[1+
(u-x)^2]}+(x^2-u^2-3)\ln{ 1+(u+x)^2\over 1+(u-x)^2}\nonumber \\ &&  +3 \ln(1+
4x^2)+{15 x\over 2} \Big[ \arctan(u+x)+\arctan(u-x)-\arctan 2x\Big] \bigg\} \,.
\end{eqnarray} 
On the other hand, one gets from the Fock diagrams with three medium
insertions:
\begin{eqnarray} 
D(k_f)^{({\rm it},F3)} &=&{9g_A^4 M_N m_\pi^2\over (4\pi f_\pi)^4 u^3}\int_0^u 
\!\!dx\Bigg\{{G \over 8} \bigg[3G -x {\partial  G \over \partial x} -u {
\partial G \over \partial u}\bigg] +{x^2\over 2}\int_{-1}^1 \!\!dy 
\int_{-1}^1  \!\! dz  \nonumber \\ && \times {yz \,\theta(y^2+z^2-1) \over 
|y z| \sqrt{y^2+z^2-1}} \bigg[ {s^2 \over 1+s^2}-\ln(1+s^2)\bigg] \Big[ \ln(1+
t^2) -t^2 \Big] \Bigg\} \,,  \end{eqnarray}
where we have introduced the auxiliary function:
\begin{equation}G(x,u) = u(1+u^2+x^2)-{1\over 4x}[1+(u+x)^2][1+(u-x)^2] \ln {
1+(u+x)^2\over 1+(u-x)^2} \,, \end{equation}
and the abbreviations $s = x y+\sqrt{u^2-x^2+x^2y^2}$ and $t = x z+\sqrt{u^2-
x^2+x^2z^2}$. Note that the expressions in eqs.(6-9) carry the large scale
enhancement factor $M_N$. It stems from the energy denominator of these
iterated diagrams which is proportional to the difference of small nucleon
kinetic energies. 
\subsection{Irreducible two-pion exchange} 
The irreducible two-pion exchange with only nucleons in the intermediate state
leads to the following contribution:  
\begin{eqnarray} 
D(k_f)^{(2\pi)}&=& {m_\pi^3\over (4\pi f_\pi)^4} \Bigg\{ \bigg[ {3\over 8u^3} 
(43g_A^4+6g_A^2-1)+{9\over 4u}(23g_A^4+2g_A^2-1)\bigg] \ln^2(u+\sqrt{1+u^2})
\nonumber \\ && +\bigg[u^2(7g_A^4-6g_A^2-1)-4-6g_A^2+46g_A^4+{3\over 4u^2} (1
-6g_A^2-43g_A^4)\bigg] \sqrt{1+u^2}\nonumber \\ && \times  \ln(u+\sqrt{1+u^2}) 
+{3  \over 8u}(43g_A^4+6g_A^2-1)+{u\over 8}(47+30g_A^2-653g_A^4)\nonumber \\
&&  +{u^3\over 4} (5+22g_A^2-19g_A^4)+u^3(15g_A^4-6g_A^2-1)\ln{m_\pi\over 
\lambda} \Bigg\} \,,  \end{eqnarray}
as obtained by differentiating eq.(14) in ref.\cite{nucmat} with respect to
$m_\pi^2$ at fixed $k_f$. We have arranged the $u^3$-terms in the last line 
such that the low-density expansion of eq.(11) starts as $k_f^3[\ln(m_\pi/
\lambda)+1/2]$ with no further additive (regularization-scheme dependent) 
constant to the chiral logarithm. Or stated differently, the underlying
$2\pi$-exchange interaction at zero momentum transfer has been restricted to 
the non-analytical piece proportional to $m_\pi^2 \ln(m_\pi/\lambda)$. The
dependence of the contribution in eq.(11) on the regularization scale 
$\lambda$ will be discussed in section 3.2.
  
\subsection{Two-pion exchange with virtual $\Delta$-isobar excitation} 
We give first the three-body contributions. The Hartree diagram (see Fig.\,2 
and eq.(5) in ref.\cite{deltamat}) leads to the following closed form  
expression:   
\begin{equation} 
D(k_f)^{(\Delta,H3)} = {3 g_A^4m^4_\pi \over \Delta(2\pi f_\pi)^4}\bigg\{u^2-
u^4+{5u^3\over 2}\arctan 2u-{1+6u^2\over 4}\ln(1+4u^2)\bigg\}\,, \end{equation}
while the contribution of the Fock diagrams can be represented as a
one-parameter integral:  
\begin{equation}
D(k_f)^{(\Delta,F3)} = {3g_A^4 m^4_\pi \over 4\Delta (4\pi f_\pi)^4 u^3}
\int_0^u \!\!dx\bigg\{2 G_S \bigg( x {\partial  G_S \over \partial x} +u 
{\partial G_S \over \partial u}-4G_S  \bigg) +G_T\bigg( x {\partial  G_T \over 
\partial x}+u {\partial G_T\over \partial u}-4G_T\bigg)\bigg\}\,,\end{equation}
with the two auxiliary functions $G_S(x,u)$ and $G_T(x,u)$ written down
explicitly in eqs.(7,8) of ref.\cite{deltamat}. As in our previous works we 
use the value $3/\sqrt{2}$ for the ratio between the $\pi N\Delta$- and $\pi 
NN$-coupling  constants.  

The two-body terms from  $2\pi$-exchange with virtual $\Delta$-excitations fall
into two classes: the dominant terms scaling reciprocally with the mass
splitting $\Delta= 293\,$MeV, and the remaining ones with a more complicated
$\Delta$-dependence. The contribution of the dominant two-body terms to the 
function $D(k_f)$ can be written again in closed form:  
\begin{eqnarray}
D(k_f)^{(\Delta2)} &=& {3\pi g_A^4m^4_\pi \over 70\Delta(2\pi f_\pi)^4}\bigg\{
( 70+14u^2 +3u^4) \arctan u \nonumber \\ && -{43+161u^2\over 
4u^3}\ln(1+u^2)   +{43 \over 4u } -{281 u \over 8} -{437 u^3 \over 6} 
\bigg\}\,. \end{eqnarray}
It has been derived from the following isoscalar central and isovector tensor
one-loop NN-scattering amplitudes \cite{2pidel}: 
\begin{equation} V_C(q) = {3g_A^4 \over 32 \pi f_\pi^4 \Delta}\bigg\{
{(2m_\pi^2 +q^2)^2 \over 2q} \arctan{q \over 2m_\pi} + m_\pi q^2 +4 m_\pi^3 
\bigg\} \,, \end{equation}
\begin{equation} W_T(q) = {g_A^4 \over 128 \pi f_\pi^4\Delta}\bigg\{
{4m_\pi^2+q^2 \over 2q} \arctan{q \over 2m_\pi}+m_\pi \bigg\}\,,\end{equation}
with $q$ the momentum transfer between the two nucleons. In these expressions 
we  have kept the polynomial pieces proportional to odd powers of the pion 
mass $m_\pi$. These non-analytic terms in the quark mass $m_q$ are a unique
feature of the chiral pion-loop dynamics. For the remaining two-body terms we 
employ the  spectral-function representation \cite{deltamat} and differentiate 
directly the imaginary parts of the $\pi N\Delta$-loop functions with respect 
to $m_\pi^2$. This gives:    
\begin{eqnarray}
D(k_f)^{(\Delta2')}&=& {3g_A^2\over (4\pi f_\pi)^4} \int_{2m_\pi}^\infty\!\!
d\mu \bigg[ 3\mu k_f -{4k_f^3 \over 3\mu} -{\mu^3 \over 2k_f} - 4 \mu^2 
\arctan{2k_f\over \mu} +{\mu^3 \over 8k_f^3}(12k_f^2+\mu^2)  \nonumber \\ &&
\times \ln\bigg( 1+{4k_f^2\over\mu^2}\bigg)\bigg] \Bigg\{\bigg[ {2 \Delta 
\over \mu} +{g_A^2 \over 8  \mu \Delta} (8 \Delta^2 +40m_\pi^2 -13 \mu^2) 
\bigg] \arctan{\sqrt{ \mu^2-4m_\pi^2}\over 2\Delta}\nonumber \\ &&-{g_A^2 \mu 
m_\pi^2 \over \Delta^2 \sqrt{\mu^2- 4m_\pi^2}}+\sqrt{\mu^2-4m_\pi^2} \bigg[ 
{3g_A^2 \mu(m_\pi^2 - \Delta^2)\over(\mu^2+ 4\Delta^2-4m_\pi^2)^2} -{2+g_A^2 
\over 2 \mu} \nonumber \\ &&+ {2g_A^2 \mu (m_\pi^2 -\Delta^2) -\mu \Delta^2 
\over 2\Delta^2(\mu^2+ 4\Delta^2-4 m_\pi^2)} \bigg]  \Bigg\}  \,,  
\end{eqnarray}   
where the $g_A^2$-terms in the curly bracket stem from box diagrams and the 
$g_A$-independent ones from the triangle diagram. The spectral-function
representation in eq.(17) involves one subtraction of a term linear in density
$\rho = 2k_f^3/3\pi^2$. The associated subtraction constant receives also 
contributions from the pion loop diagrams with a non-analytical dependence on 
the quark mass. We reinstore these distinguished pieces from the chiral 
pion-loop dynamics by the additional contribution linear in density:       
\begin{equation} D(k_f)^{(dt)} = {3 g_A^2 k_f^3\over (4\pi f_\pi)^4} \Bigg\{ 
(2-5g_A^2) \ln {m_\pi \over 2\Delta} +{4\Delta^2 -5g_A^2(2\Delta^2+3 m_\pi^2)
\over 2\Delta \sqrt{\Delta^2-m_\pi^2} }\ln {\Delta+  \sqrt{\Delta^2-m_\pi^2}
\over m_\pi} \Bigg\} \,, \end{equation}
which has the property that it vanishes in the chiral limit. 

\subsection{Chiral $\pi\pi NN$ contact vertex proportional to $c_1$}
\begin{figure}
\begin{center}
\includegraphics[scale=0.8,clip]{c1fig.epsi}
\end{center}
\vspace{-0.2cm}
{\it Fig.\,2: Hartree and Fock diagrams related to
the chiral $\pi\pi NN$ contact vertex proportional to $c_1$. The combinatoric 
factors of these diagrams are 1/2 and 1, in the order shown.}
\end{figure}

In chiral perturbation theory the nucleon sigma-term $\sigma_N$ has a specific
nonlinear dependence on the quark mass $m_q$. The leading linear term comes
from the effective Lagrangian ${\cal L}_{\pi N}^{(2)} = c_1\,\bar N N {\rm tr} 
\chi_+=4c_1 m_\pi^2 \bar NN\sqrt{1-\vec \pi^{\,2}/f_\pi^2}$ and the nonlinearities 
arise from pion-loops and higher order counterterms. Putting together all 
(numerically) relevant pieces the ratio $\sigma_N/m_\pi^2$ entering eq.(1)
takes the form:       
\begin{eqnarray} {\sigma_N \over m_\pi^2} &=& -4c_1 -{9 g_A^2 m_\pi \over 64\pi
f_\pi^2} + {3c_1 m_\pi^2 \over 2\pi^2 f_\pi^2}\ln{m_\pi \over \lambda}
\nonumber\\ &&  + {9g_A^2  \over (4\pi f_\pi)^2}  \Bigg\{ \Delta \ln 
{m_\pi \over2\Delta} +\sqrt{\Delta^2-m_\pi^2} \ln {\Delta+  \sqrt{\Delta^2-
m_\pi^2} \over m_\pi} \Bigg\} \,. \end{eqnarray}
By chiral symmetry, the $c_1$-term in the effective Lagrangian generates also 
an additional $\pi\pi NN$-contact interaction with vertex insertion: $-4ic_1 
m_\pi^2 f_\pi^{-2} \delta^{ab}$. On the one hand it makes up the sizeable 
logarithmic loop correction to $\sigma_N/m_\pi^2$ in eq.(19). On the other 
hand it gives rise to an additional two-pion exchange contribution to the
NN-interaction, and moreover it generates a long-range three-nucleon force
(see diagrams in Fig.\,2). The two-body terms in nuclear matter lead to the 
following contribution to the $m_\pi^2$-derivative of the energy per particle:   
\begin{equation}
D(k_f)^{(c_1,2)} = {3g_A^2 c_1m^4_\pi \over 280\pi^3 f_\pi^4}\bigg\{
(14u^2 +3u^4) \arctan u +{27+49u^2\over 4u^3}\ln(1+u^2)   -{27 \over 4u } 
-{71 u \over 8} -{99u^3 \over 2} \bigg\}\,, \end{equation}
which has been derived from the isoscalar central one-loop NN-scattering 
amplitude:
\begin{equation} V_C(q) = {3g_A^2 c_1 m_\pi^2 \over 4 \pi f_\pi^4}\bigg\{
{2m_\pi^2 +q^2 \over 2q} \arctan{q \over 2m_\pi} + m_\pi\bigg\} \,.
\end{equation}
In addition, there are $c_1$-contributions from the three-body Hartree 
diagram: 
\begin{equation}
D(k_f)^{(c_1,H3)} = {3g_A^2 c_1m^4_\pi \over (2\pi f_\pi)^4}\bigg\{
3u^2 -2u^4+6u^3 \arctan 2u -\bigg({3\over 4}+4u^2 \bigg) \ln(1+4u^2)
\bigg\}\,, \end{equation} 
and the three-body Fock diagram:
\begin{equation}
D(k_f)^{(c_1,F3)} = {9g_A^2 c_1m^4_\pi \over (4\pi f_\pi)^4 u^3} \int_0^u 
\!\!\!dx \, G \bigg[ 4  G -x {\partial  G \over \partial x} -u {\partial G 
\over \partial u}  \bigg] \,,  \end{equation} 
with the auxiliary function $G(x,u)$ defined in eq.(10). 
\subsection{Chiral limit}
The $m_\pi^2$-derivative of the energy per particle, $D(k_f)$, presented in
the previous subsections takes a particularly simple form in the chiral limit 
$m_\pi=0$. In that limiting case almost all integrals can be solved and the 
dependence  on the Fermi momentum $k_f$ becomes simply powerlike (with an 
exponent determined by the mass dimension of the prefactor). The subscript $0$ 
denotes the function $D(k_f)$ in the chiral limit $m_\pi \to 0$. The 
$1\pi$-exchange Fock term (with the renormalization factor in the chiral limit,
$\Gamma(0)=1$) gives:
\begin{equation} D_0(k_f)^{(1\pi)} = {g_A^2 k_f \over(4\pi f_\pi)^2} \bigg( 
{k_f^2 \over M_N^2}-{9\over    4}\bigg)  \,, \end{equation}
and the total contribution from iterated $1\pi$-exchange reads:
\begin{equation} D_0(k_f)^{(\rm it)} = {g_A^4M_N k_f^2\over 5(4\pi f_\pi)^4} 
\Big(8\pi^2+36  \ln 2-3\Big)  \,. \end{equation}
The contribution from irreducible $2\pi$-exchange:
\begin{equation} D_0(k_f)^{(2\pi)} = {k_f^3 \over (4\pi f_\pi)^4} 
\bigg\{(g_A^2-1)(7g_A^2 +1) \ln {2k_f \over \lambda} +{1\over 4}(5+22g_A^2-
19g_A^4)+8 g_A^4 \ln{m_\pi \over \lambda}\bigg\}\,, \end{equation}
has a logarithmic singularity $\ln(m_\pi/\lambda)$. It gets exactly canceled
by the $m_\pi^2$-derivative of the renormalization factor $\Gamma(m_\pi)$ applied
to the static $1\pi$-exchange: 
\begin{equation} D_0(k_f)^{(\rm ren)} = {k_f^3 \over (2\pi f_\pi)^4} \bigg\{ 
g_A^4 \bigg(\gamma- {1\over 2}\ln{m_\pi \over \lambda}\bigg)+C \bigg\}\,.  
\end{equation}
This crucial feature instructs us that one must work consistently with the 
parameters in the chiral limit as they are given by the effective chiral 
Lagrangian. It is mandatory to include only those renormalization effects to 
physical parameters which are actually generated by the pion-loops to the 
order one is working. The constant $C$ represents the additional effect of a 
NN-contact coupling linear in the quark mass  $m_q$, and $\gamma$ is a 
left-over from the Goldberger-Treiman discrepancy. The two- and three-body 
terms from $2\pi$-exchange with single $\Delta$-excitation scaling as 
$1/\Delta$ read together:
\begin{equation} D_0(k_f)^{(\Delta 3)} = {g_A^4 k_f^4 \over \Delta 
(4\pi f_\pi)^4} \bigg( {12\pi^2 \over 35} -25 \bigg)  \,, \end{equation}
where the two-body term eq.(14) has contributed $36\pi^2/35$ to the numerical 
factor in brackets. For the remaining two-body Fock terms from $2\pi$-exchange 
with $\Delta$-excitations the spectral-function representation turns into:  
\begin{eqnarray} D_0(k_f)^{(\Delta 2')}&=& {g_A^2\Delta^3\over (4\pi f_\pi)^4} 
\int_0^\infty \!\! dx\,x\,\bigg\{g_A^2 \bigg[{2x+11x^3+ 6x^5\over (1+x^2)^2} 
+(13x^2-2) \arctan x\bigg] \nonumber\\ && \qquad \qquad\qquad \qquad +{4x+6x^3 
\over 1+x^2} -4\arctan x  \bigg\}\,\Phi\bigg({k_f \over x \Delta}\bigg) \,,
\end{eqnarray} 
with the auxiliary function:
\begin{equation} \Phi(y) = {6\over y}-9 y +y^3+24 \arctan y -{6\over y^3}
(1+3y^2) \ln(1+y^2) \,.\end{equation}
For low densities the contribution in eq.(29) behaves as $k_f^5\ln(k_f/\Delta 
)$. Finally, the total contribution from the $\pi\pi NN$-contact vertex 
proportional to $c_1$ reads, in the chiral limit:
\begin{equation} D_0(k_f)^{(c_1)} = {3g_A^2 c_1 k_f^4 \over (2\pi f_\pi)^4} 
\bigg( {\pi^2 \over 35} -{5\over 4} \bigg)  \,. \end{equation}
\section{Results}
This section presents and discusses results for the in-medium chiral
condensate as a function of baryon density $\rho$. Apart from just collecting
and computing the series of terms given in section 2, this includes also a
detailed investigation of the pion mass dependence of the quark condensate at
given density $\rho$. It is furthermore necessary to estimate the quark mass
dependence of the NN-contact term (i.e. the size of the parameter $C$
introduced in eq.(27)) which encodes short-distance dynamics not controlled by
the underlying chiral effective field theory. At this point we can now take
very recent computations of the NN-potential from lattice QCD for orientation. 

\begin{figure}
\begin{center}
\includegraphics[scale=0.55,clip]{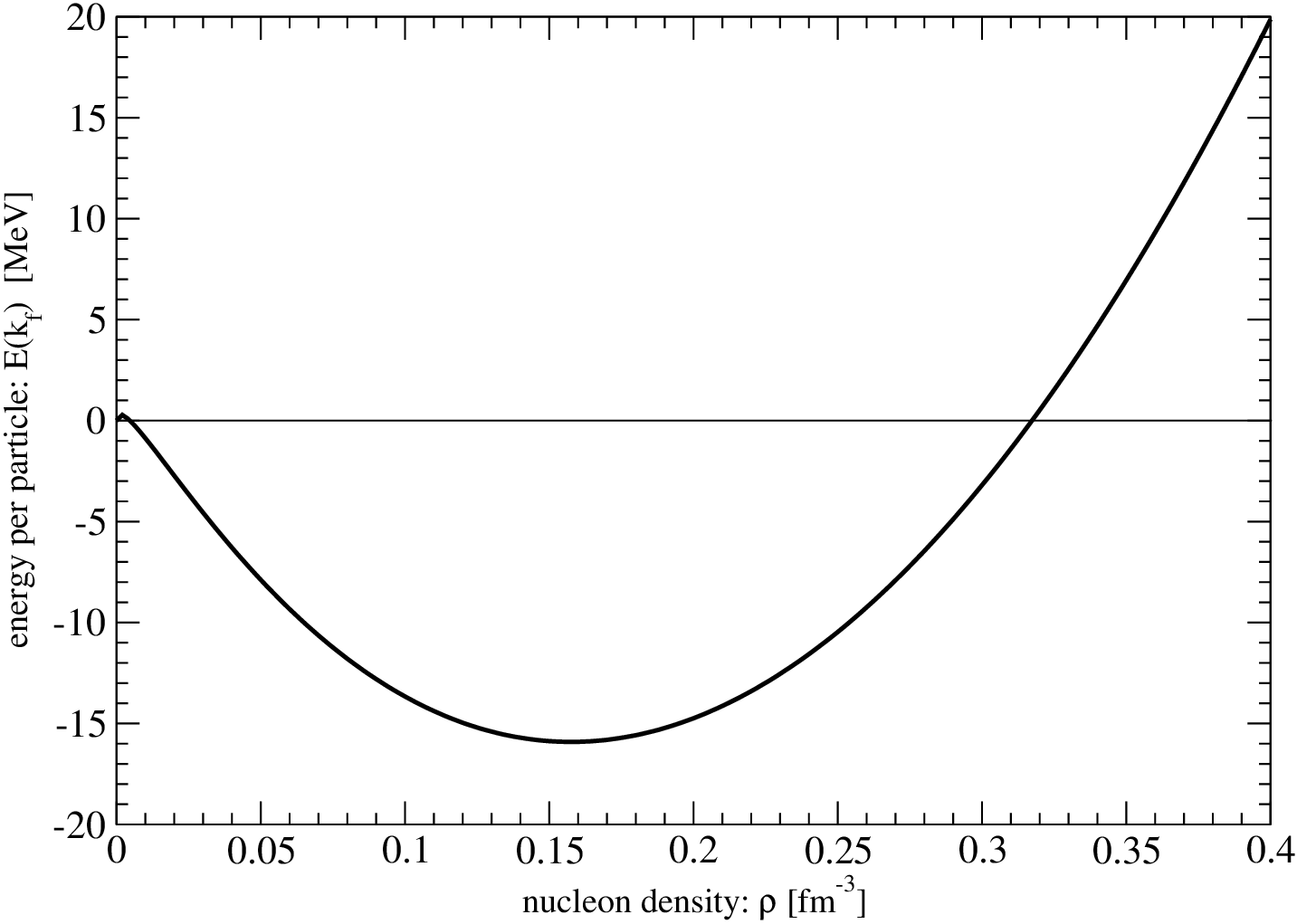}
\end{center}
\vspace{-0.2cm}
{\it Fig.\,3: The nuclear matter saturation curve underlying our calculation
of the in-medium quark condensate. Apart from the interaction contributions 
described in refs.\cite{nucmat,deltamat} and those proportional to $c_1$, it 
includes one single adjusted term linear in density, $\bar E(k_f)^{({\rm adj})} = -
7.64\,{\rm GeV}^{-2}\, k_f^3$.}   
\end{figure}

\subsection{Parameter fixing}
First, we have to fix the parameters. The pion decay constant in the chiral
limit $f_\pi$ is determined by the relation: $f_{\pi,{\rm phys}} = f_\pi [1+
\bar l_4 (m_\pi/4\pi f_\pi)^2]=92.4\,$MeV. Choosing the central value $\bar 
l_4 = 4.4\pm 0.2$ of ref.\cite{cola} one gets $f_\pi = 86.5\,$MeV. For the 
nucleon axial vector coupling constant $g_A$ in the chiral limit we take the 
value $g_A=1.224$ as obtained recently via chiral extrapolations of lattice 
data in ref.\cite{musch}. A similar analysis \cite{massi} gives for the 
nucleon mass in the chiral limit $M_N = 882\,$MeV and for the low-energy 
constant $c_1 = -0.93\,$GeV$^{-1}$ (as central values). For the delta-nucleon 
mass splitting we take the empirical value $\Delta = 293\,$MeV. This is 
consistent to the order in the loop expansion we are working here. The 
parameter $\gamma$ introduced in eq.(3) is determined by the relation: $(g_{\pi 
N}/M_N)_{\rm phys}=\sqrt{\Gamma(m_\pi)}\, g_A/f_\pi$. Taking for the left hand 
side $13.2/(939\,$MeV), we deduce $\gamma=-1.505$ at the regularization scale 
$\lambda =M_N =882\,$MeV. Or stated differently, for our choice of parameters 
($g_A$ and $f_\pi$ in the chiral limit) the Goldberger-Treiman relation is 
exact within one percent. We neglect the $1.2\%$-difference between the
physical pion mass $m_{\pi,{\rm  phys}} = m_\pi [1- \bar l_3 (m_\pi/8\pi f_\pi)^2]$ 
(with $\bar l_3 \simeq 3$ \cite{cola}) and the leading order one, $m_\pi$, since 
this difference is much smaller than the splitting between charged and neutral
pion masses.  

With these fixed parameters we obtain (for the physical value of the pion mass 
$m_\pi = 135\,$MeV) the nuclear matter equation of state as shown in Fig.\,3. 
Besides the $1\pi$- and $2\pi$-exchange contributions described in
refs.\cite{nucmat,deltamat} and those proportional to $c_1$, it includes one
single adjusted term linear in density: $\bar E(k_f)^{({\rm adj})}=-7.64\,{\rm GeV
}^{-2}\,k_f^3$. We interpret its strength parameter to subsume all unresolved
short-distance NN-dynamics relevant for nuclear matter at low and moderate
densities $0 \leq \rho \leq 2.5\rho_0=0.4\,$fm$^{-3}$. Its (weak) implicit
quark mass dependence will be estimated below. The nuclear matter 
compressibility $K = k_{f0}^2 \bar E''(k_{f0})$ related to the curvature of the 
saturation curve at its minimum comes out as $K= 295\,$MeV. This value lies at
the high side of present semi-empirical determinations of $K$ \cite{dario}.  
    
\begin{figure}
\begin{center}
\includegraphics[scale=0.55,clip]{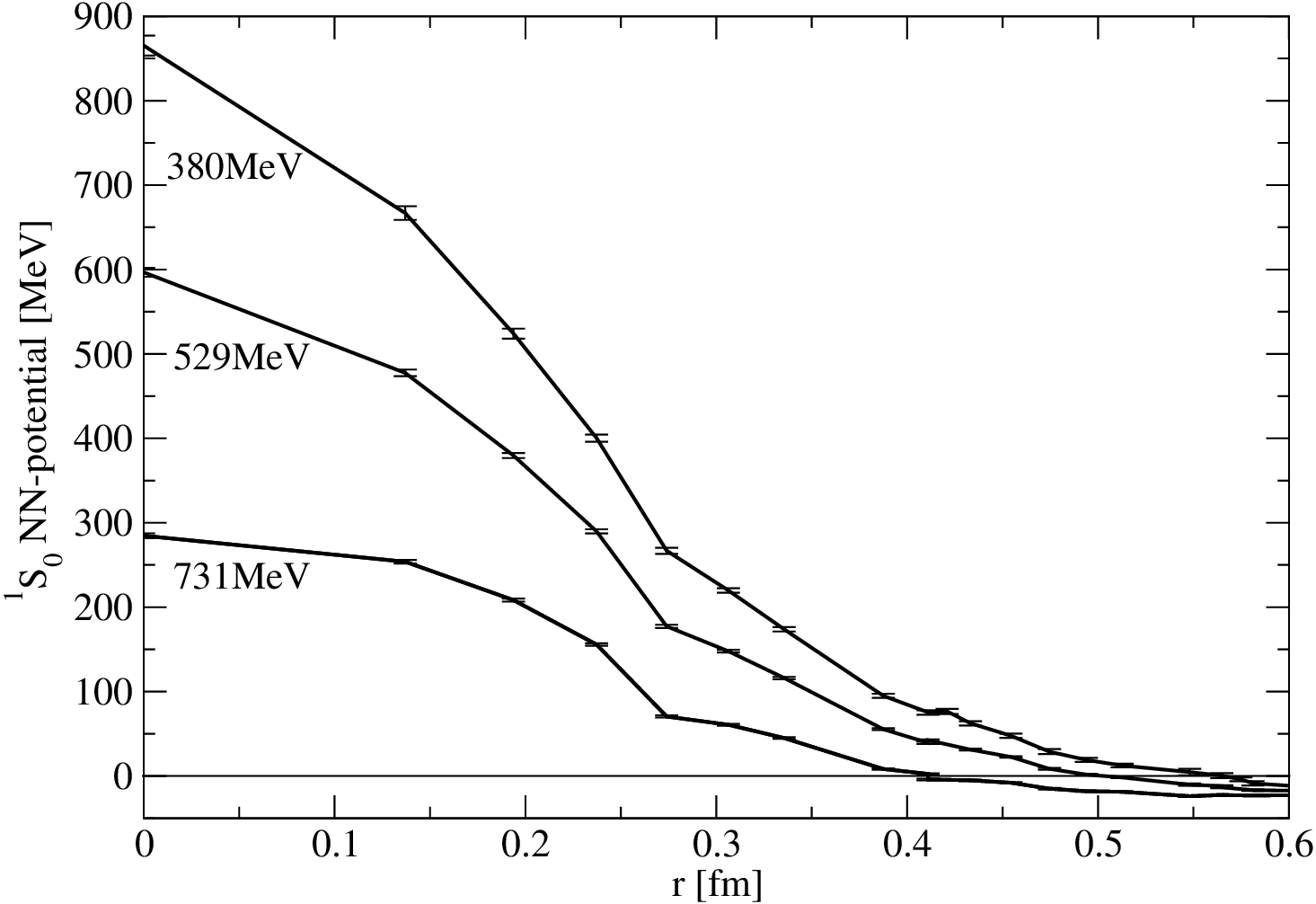}
\end{center}
\vspace{-0.5cm}
{\it Fig.\,4: The nucleon-nucleon potential in the $^1S_0$ channel from
lattice QCD \cite{hatsuda} for three different pion masses, $m_\pi = 
(380,\,529,\,731)\,$MeV.} 
\end{figure}

Next, we estimate the parameter $C$ introduced in eq.(27) which represents 
quark mass dependent effects from the short-range NN-interaction. The simple 
model of $\omega(782)$-meson exchange would give $C^{(\omega)} \simeq -0.7$, a
large correction, choosing an $\omega N$-coupling constant of order $10$ and 
the constituent quark counting rule $\partial m_\omega/\partial m_q=2$. However,
since the $\omega$-exchange model has no clear connection to the short-range 
NN-dynamics of QCD, we follow another option. Recently, the nucleon-nucleon
potential has been studied within lattice QCD \cite{hatsuda} using the
quenched approximation. The short-distance part ($r \leq 0.6\,$fm) of this
potential in the $^1S_0$ channel is shown in Fig.\,4 for three different pion
masses, $m_\pi =(380,\,529,\,731)\,$MeV. We can identify the volume integrals 
$I_0=4\pi\int_0^{r_0} dr\,r^2V(r)$ over the repulsive cores of these potentials 
with the strength of a contact-coupling in the $^1S_0$ channel. From the three 
values $I_0= (83.6,\,53.3, \,21.9)\,$MeVfm$^3$ we obtain a mean value for the 
derivative with respect to the squared pion mass: $\partial I_0/\partial 
m_\pi^2\simeq -0.17 \,$GeV$^{-1}$fm$^3$. Via the contribution $\bar E(k_f)^
{(\rm sd)}=I_0\,k_f^3/4\pi^2$ to the energy per particle we estimate the 
parameter $C$ as $C = 4\pi^2 f_\pi^4 \,\partial I_0/\partial m_\pi^2  \simeq 
-0.05$. In comparison to the $\omega$-meson exchange this is a rather small 
number. These considerations raise also some doubts concerning the 
significance of the vector boson exchange phenomenology as a valid picture of 
the short-distance NN-dynamics. Even with a factor 2, to include an equally 
strong contribution from the $^3S_1$ channel \cite{hatsuda}, the value $C 
\simeq -0.1$ affects the condensate ratio at nuclear matter saturation 
density $\rho_0=0.16\,$fm$^{-3}$ (corresponding to $k_{f0} = 263\,$MeV) only at 
the 3 permille level (and 4 times as much at $2\rho_0$). 

We can therefore conclude that the short-range NN-dynamics as given by lattice 
QCD \cite{hatsuda} has a negligible influence on the in-medium chiral 
condensate $\langle \bar qq \rangle(\rho)$. The deviations from the linear 
density approximation are primarily caused by the long- and intermediate range
$1\pi$- and $2\pi$-exchange dynamics. 

There is some residual dependence on the regularization scale $\lambda$ left 
over which is not balanced by the parameters $l_3^r(\lambda)$ and
$\gamma(\lambda)$ (namely from the last term in eq.(11)).\footnote{In
principle, this scale dependence is balanced by the parameter $C(\lambda)$ in
eq.(27). But this (formal) point of view introduces the need to fix the scale
$\lambda$ in an estimate of $C$.} Varying $\lambda$ between $0.6\,$GeV and 
$1.2\,$GeV changes the condensate ratio at $\rho_0$ by $3.5\%$. Since this is
much smaller than the effect induced by the uncertainty of the empirical
nucleon sigma-term $\sigma_N = (45\pm 8)\,$MeV \cite{sigma} we stay with the 
''natural'' choice of $\lambda = M_N=882\,$MeV. When inserting into eq.(19) it 
reproduces correctly $\sigma_N= 44.3 \,$MeV for $m_\pi=135\,$MeV. 
\subsection{In-medium condensate}
\begin{figure}
\begin{center}
\includegraphics[scale=0.55,clip]{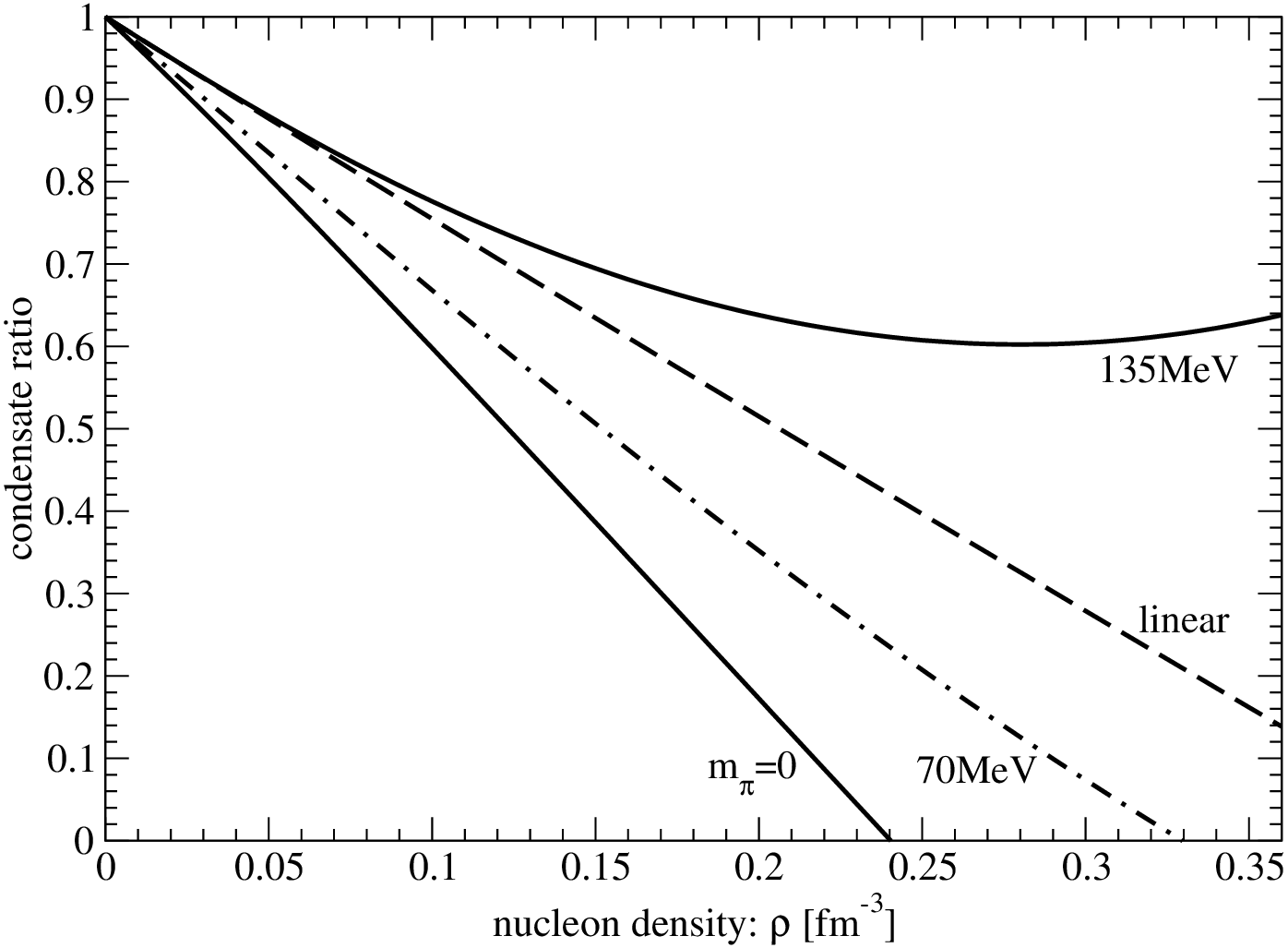}
\end{center}
\vspace{-0.2cm}
{\it Fig.\,5: The ratio of the in-medium chiral condensate to its vacuum value 
as a function of the nucleon density $\rho$ for three different values of the
pion mass, $m_\pi = (0,\,70,\, 135)\,$MeV. The dashed line corresponds to the 
linear density approximation using the empirical central value $\sigma_N = 
45\,$MeV \cite{sigma}.} 
\end{figure}

We are now in the position to present and discuss numerical results for the
in-medium quark condensate. Fig.\,5 shows the condensate ratio $\langle \bar 
qq \rangle(\rho)/\langle 0|\bar qq|0 \rangle$ as a function of the nucleon 
density $\rho = 2k_f^3/3\pi^2$ in the region $0 \leq \rho \leq 0.36\,$fm$^{-3}
=2.25\rho_0$ (i.e. $k_f\leq 345\,$MeV) for three different values of the pion 
mass, $m_\pi =(0,\,70,\, 135)\,$MeV. The dashed line in Fig.\,5 corresponds to 
the linear density approximation using the empirical central value $\sigma_N 
= 45\,$MeV of the nucleon sigma-term. One observes a very strong and nonlinear 
dependence of the ''dropping'' condensate on the actual value of the pion mass
$m_\pi$. In the chiral limit, $m_\pi = 0$, the quark condensate decreases 
effectively with a slope $1.8$ times as large as given by the linear density 
approximation. As a consequence chiral symmetry would be restored already at 
about $1.5\rho_0$ if the up- and down-quark masses were strictly zero. This 
faster decrease is caused by two features: First, the ratio $\sigma_N/m_\pi^2=
-4c_1 =3.72\,$GeV$^{-1}$ is, in the chiral limit, about 1.5 times larger than
at the physical point, $45/135^2\,$MeV$^{-1} = 2.47\,$GeV$^{-1}$. Secondly, the 
two-pion exchange effects in the chiral limit drive the condensate ratio 
further down. Of course, the actual density at which chiral symmetry 
restoration would occur can only be roughly estimated in our calculation: once 
the chiral condensate becomes too small the very foundation of the chiral 
effective field theory approach to nuclear matter (namely the spontaneous 
breaking of chiral symmetry in the vacuum) is lost. 

At the physical value of the pion mass, $m_\pi = 135\,$MeV, the density
dependence of the condensate ratio $\langle \bar qq \rangle(\rho)/\langle 0|
\bar qq|0\rangle$ is drastically different. At densities around $1.8\rho_0$ 
the in-medium condensate stabilizes now at about $60\%$ of its vacuum value, 
and there is no further reduction in the entirely density region where
the present chiral approach to nuclear matter can be trusted. For higher
values of the pion mass the effects counteracting chiral restoration become
even more pronounced.

\begin{figure}
\begin{center}
\includegraphics[scale=0.55,clip]{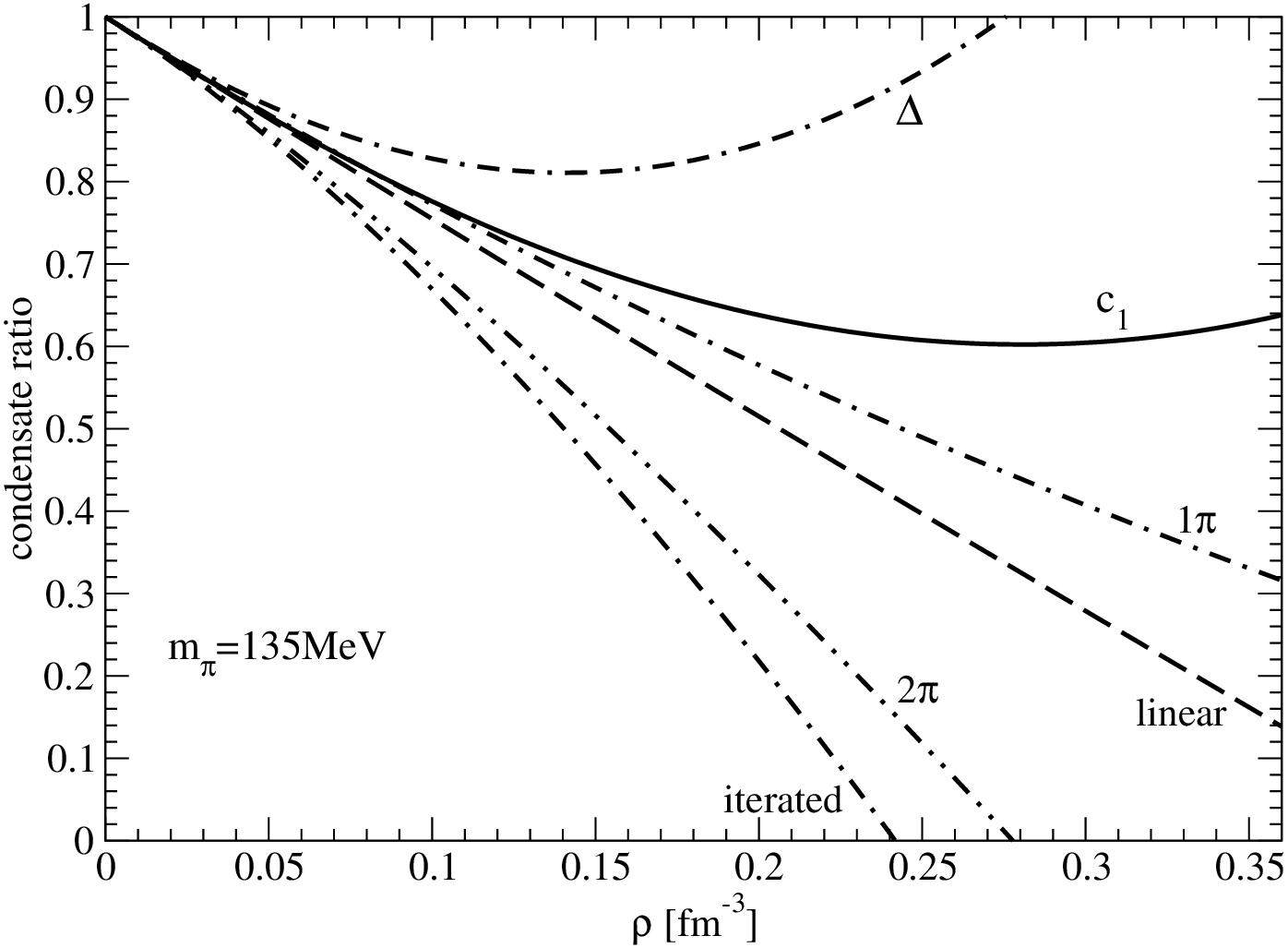}
\end{center}
\vspace{-0.2cm}
{\it Fig.\,6: Interaction contributions to the ratio between in-medium and
vacuum chiral condensate. The five classes described in subsections 2.1-2.5 
are  consecutively added in the sequence: $linear \to 1\pi \to iterated \to 
2\pi \to \Delta \to c_1$.}
\end{figure}

Let us have a closer look at individual contributions. At $2\rho_0= 0.32\,
$fm$^{-3}$ (corresponding to $k_f = 331.4\,$MeV) one gets from the sequence of
the five classes of interaction contributions (described in subsections 2.1 to 
2.5) a total correction to $\langle \bar qq \rangle(\rho)/\langle 0| \bar qq|0
\rangle$ beyond the linear density approximation of $(0.14-0.83 +0.27 +1.34-
0.54)=0.38$. Notice the cancellation between large contributions of opposite 
signs from iterated $1\pi$-exchange and $2\pi$-exchange with virtual $\Delta(
1232)$-excitation. At normal nuclear matter density $\rho_0=0.16\,$fm$^{-3}$ 
the individual entries are about a factor 4 smaller: $(0.04-0.24+0.07+0.33
-0.13) = 0.07$, suggesting an approximate $\rho^2$-dependence of the total 
interaction contribution. 

Fig.\,6 shows separately the effects of the five classes of interaction
contributions. They are consecutively added up in the sequence: $linear \to
1\pi \to iterated \to 2\pi \to \Delta \to c_1$. The last two major steps (in 
opposite directions) should not be misinterpreted as a sign of bad convergence 
since the corresponding terms belong to the same order in the small momentum 
expansion  of the nuclear matter energy density ($-4c_1\simeq 3g_A^2/4\Delta$). 
Fig.\,6 actually demonstrates the prominent importance of the $2\pi$-exchange 
interaction beyond leading order for the in-medium quark condensate. 

It is important to include all effects generated by the pion-loops. For 
example, if one would drop the last constant term proportional to $4m_\pi^3 
\sim m_q^{3/2}$ in eq.(15) an amount of $0.25$ would be missing in the total 
balance (at $\rho_0$). As a consequence of that omission the condensate ratio 
would lie appreciably below the linear density approximation. This particular 
non-analytical $4m_\pi^3$-term gives also an explanation for the drastically 
different behavior of the in-medium condensate in the chiral limit and for the 
physical pion mass. Its contribution to the condensate ratio, $27 g_A^4 m_\pi 
\rho^2/(128 \pi f_\pi^6\Delta)$, vanishes in the chiral limit, $m_\pi = 0$, but 
reaches $100\%$ for the physical pion mass $m_\pi = 135\,$MeV at $\rho =0.32\,
$fm$^{-3} = 2 \rho_0$. This selective consideration does, of course, not mean
that the other one- and two-pion exchange contributions would not also depend 
strongly on the pion mass. Their dependences can be studied easily case by 
case with the help of the analytical formulas presented in section 2. In 
general, one can say that the condensate ratio $\langle \bar qq \rangle(\rho)/
\langle 0|\bar qq|0 \rangle$ is affected significantly by interaction terms 
which otherwise play only a marginal role for the nuclear matter equation of 
state $\bar E(k_f)$ (as e.g. the chiral $\pi\pi NN$-contact term proportional
to $c_1$). This pronounced shifting of weights comes from taking the
derivative with respect to the squared pion mass $m_\pi^2$. On the other hand, 
the Fermi gas approximation (i.e. the linear density approximation) works
reasonably well for the in-medium chiral condensate almost up to nuclear
matter saturation density $\rho_0=0.16\,$fm$^{-3}$, even though it is far from
correctly describing nuclear matter as a self-bound Fermi liquid.  

Let us compare our results for the in-medium chiral condensate with other
works which have treated to some limited extent the pion-exchange dynamics in 
nuclear matter. In the work of Lutz et al.\,\cite{lutz} $1\pi$-exchange plus 
an adjustable  NN-contact interaction have been iterated to second order. In 
the region $0 \leq \rho \leq 2\rho_0$ they find a  weaker (positive) 
deviation from the linear density approximation, with no trend for 
stabilization of the in-medium condensate. This difference comes from not
including the irreducible $2\pi$-exchange, the chiral $\pi N\Delta$-dynamics
and the $c_1$-contact vertex. Recently, the T\"ubingen group \cite{fuchs} has
employed the chiral nucleon-nucleon potential at next-to-leading order to
calculate nuclear matter within the relativistic Dirac-Brueckner-Hartree-Fock 
approach. They also derive the in-medium chiral condensate by exploiting the 
Feynman-Hellmann theorem eq.(1). Irrespective of using the Hartree-Fock or 
Brueckner-Hartree-Fock approximation, their effects from pion-exchange 
interactions (e.g. $\sim 15\%$ at $2\rho_0$ \cite{fuchs}) are much smaller
than in our calculation. Moreover, there is no trend of stabilization in the 
whole density region $0 \leq \rho \leq 3 \rho_0$ considered. Again, these 
substantial differences arise in ref.\cite{fuchs} from taking only the 
next-to-leading order chiral NN-potential (i.e. $1\pi$-exchange and 
irreducible $2\pi$-exchange), but neglecting the actually more important 
effects from $2\pi$-exchange with virtual $\Delta$-isobar excitation. We note
that our perturbative calculation, when truncated to $1\pi$-exchange, iterated
$1\pi$-exchange and irreducible $2\pi$-exchange exchange, would lead to an 
in-medium condensate below the linear density approximation (see Fig.\,6). 
Qualitative differences may also come from the approximations (angle-averaged 
Pauli-blocking operator, etc.) used in the Brueckner-Hartree-Fock calculation 
of ref.\cite{fuchs} and not treating the iterated $1\pi$-exchange in full 
detail (as done in the present work). Apart from all these differences, 
ref.\cite{fuchs} agrees with our conclusion that the short-range NN-dynamics 
plays only a minor role for the in-medium chiral condensate. It is indicated 
to repeat the calculation of ref.\cite{fuchs} with the more
complete next-to-next-to-leading order chiral NN-potential including the
non-analytical polynomial pieces.

Our results for the in-medium chiral condensate can be alternatively
summarized by the density dependent effective nucleon sigma-term:
\begin{equation} \sigma_{N,{\rm eff}}(\rho) = \sigma_N \bigg(1 -{3k_f^2 \over
10 M_N^2} +{9k_f^4 \over 56 M_N^4}   \bigg) + m_\pi^2 D(k_f) \,. \end{equation} 
It captures the correlation effects in the nuclear medium which reduce the 
tendency towards chiral symmetry restoration. Fig.\,7 shows the effective 
nucleon sigma-term\footnote{A qualitatively similar result, though based on a
different approach, has been reported in ref.\cite{bentz}.} as a function of 
the density for $0 \leq \rho \leq  0.36\,$fm$^{-3}$. In this figure we display 
also the uncertainties associated with the still existing error band of $\pm 
8\,$MeV in the empirical determination of $\sigma_N$. Moreover, we have varied 
the regularization scale $\lambda$ between $0.6\,$GeV and $1.2\,$GeV which 
leads to some widening of the error band as the density increases.        
\begin{figure}
\begin{center}
\includegraphics[scale=0.55,clip]{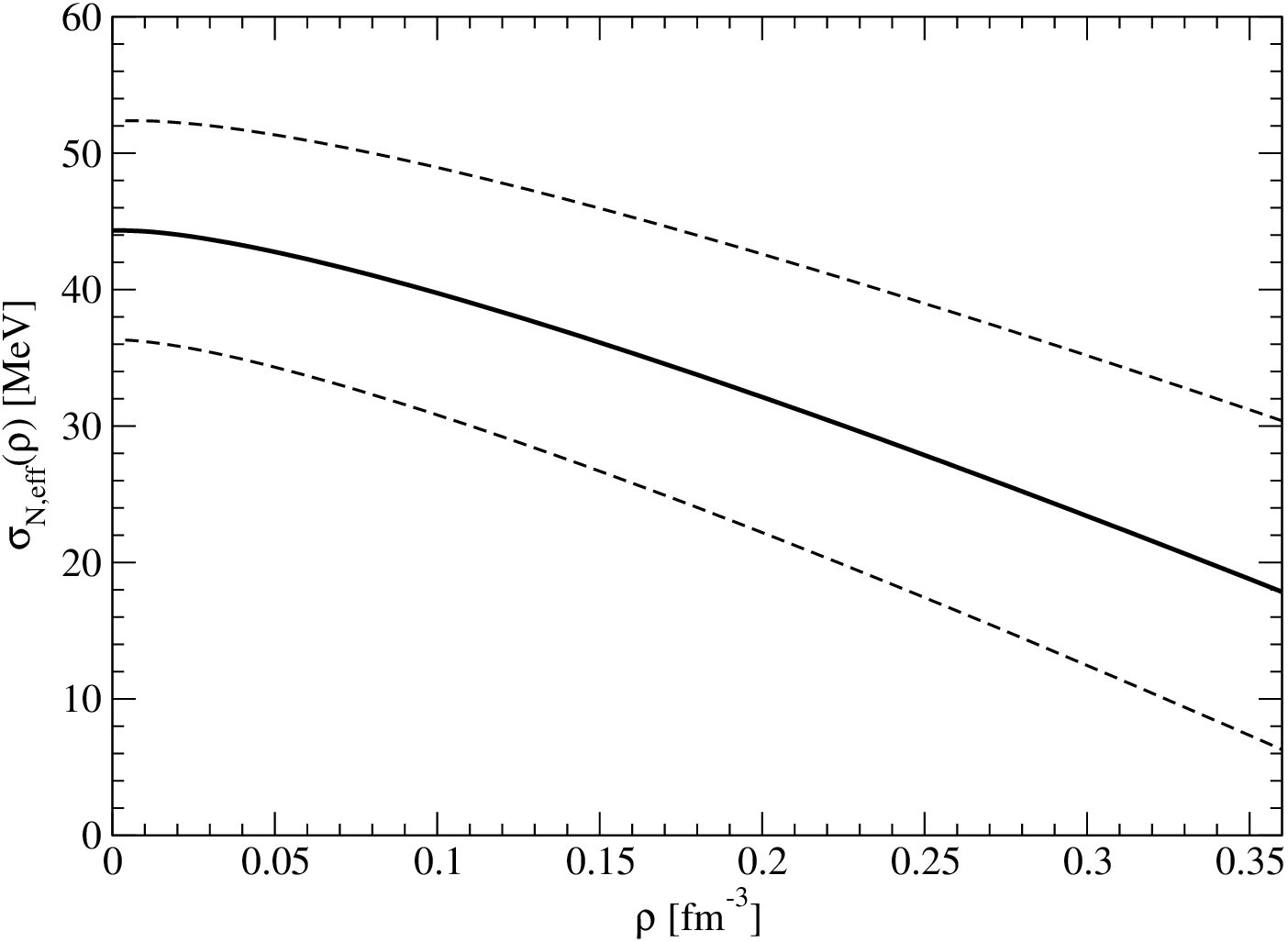}
\end{center}
\vspace{-0.2cm}
{\it Fig.\,7: The effective in-medium  nucleon sigma-term $\sigma_{N,{\rm
eff}}(\rho)$ versus the nucleon density $\rho$ with its error band $\pm 8\,$MeV.}
\end{figure}

\section{Summary and concluding remarks} 
In this work we have used in-medium chiral perturbation theory to calculate 
the quark condensate $\langle \bar qq \rangle(\rho)$ beyond the linear density 
approximation. The pertinent correction term follows from differentiating the 
interaction contributions to the energy per particle of isospin-symmetric 
nuclear matter with respect to the pion mass. Analytical expressions for the 
contributions to $D(k_f)=\partial \bar E(k_f)/\partial m_\pi^2$ from $1\pi
$-exchange (with $m_\pi$-dependent vertex corrections), iterated $1\pi
$-exchange and irreducible $2\pi$-exchange with inclusion of $\Delta$-isobar 
excitations and Pauli-blocking corrections have been presented in section 2. 

We find a strong, nonlinear dependence of the ''dropping'' in-medium 
condensate on the value of the pion (or light quark) mass. In the chiral 
limit, $m_\pi=0$, chiral restoration seems to be reached already at about $1.5$ 
times normal nuclear matter density. By contrast, for the physical pion mass 
$m_\pi = 135\,$MeV, the in-medium condensate stabilizes at about $60\%$ of its 
vacuum value above that same density. Including systematically the effects 
from $2\pi$-exchange with $\Delta(1232)$-isobar excitation (or the equivalent 
chiral $\pi\pi NN$ contact interactions $c_{2,3,4}$) is crucial in order to 
obtain such a pronounced behavior. Non-analytical (contact) terms in the quark 
mass (such as $m_\pi^3\sim m_q^{3/2}$), which are often dropped in the 
presentation of the chiral NN-potential, have a very strong influence on the 
in-medium chiral condensate. Below $3\rho_0/4= 0.12\, $fm$^{-3}$ the correction 
beyond the linear density  approximation remain relatively small. This finding 
can be taken as an a posteriori justification of the assumptions made in
ref.\cite{finelli} about the in-medium scalar mean-field. 

As a consequence of the hindered tendency towards chiral symmetry restoration
(in the real world with $m_\pi=135\,$MeV), pions and nucleons can be used as
effective low-energy  degrees of freedom at least up to twice nuclear matter 
density. The major source of uncertainty for the density dependence of 
$\langle \bar qq \rangle(\rho)$ is caused by the error band in the empirical 
determination of the nucleon sigma-term $\sigma_N=(45\pm 8)\,$MeV. One can
hope that upcoming dispersion relation analyses of $\pi N$-scattering data and
lattice QCD calculations will lead to a more accurate value of $\sigma_N$. Of
course, there remain also questions about the size of effects from yet higher 
order interaction contributions related to $3\pi$-exchange, two-loop 
$2\pi$-exchange etc. On the other hand, an estimate based on recent lattice
QCD results indicates that the short-distance NN-dynamics has only a very small
effect on the density dependence of the quark condensate $\langle \bar qq
\rangle(\rho)$.       

\section*{Acknowledgments} We thank T. Hatsuda and N. Ishii for providing 
us with the nucleon-nucleon potential from lattice QCD and for stimulating
discussions.    


\vspace{-0.3cm}
\begin{thebibliography}{99}
\bibitem{gerber} P. Gerber and H. Leutwyler, {\it Nucl. Phys.} {\bf B321}, 387 
(1989).\vs
\bibitem{cheng} M. Cheng et al., {\it Phys. Rev.} {\bf D74}, 054507 (2006);
hep-lat/0710.0354.\vs 
\bibitem{aoki} Y. Aoki, Z. Fodor, S.D. Katz, and K.K. Szabo, {\it Phys. Lett.} 
{\bf B643}, 46 (2006).\vs
\bibitem{sigma} J. Gasser, H. Leutwyler and M.E. Sainio, {\it Phys. Lett.}
{\bf B253}, 252 (1991).\vs
\bibitem{cohen} T.D. Cohen, R.J. Furnstahl, and D.K. Griegel, {\it Phys. Rev.} 
{\bf C45}, 1881 (1992).\vs
\bibitem{ko} G.Q. Li and C.M. Ko, {\it Phys. Lett.} {\bf B338}, 118 
(1994).\vs
\bibitem{rolf} R. Brockmann and W. Weise, {\it Phys. Lett.} {\bf B367}, 40 
(1996).\vs
\bibitem{lutz} M. Lutz, B. Friman and C. Appel, {\it Phys. Lett.} {\bf B474},
 7 (2000).\vs
\bibitem{nucmat} N. Kaiser, S. Fritsch and W. Weise, {\it Nucl. Phys.} {\bf
A697}, 255 (2002).\vs
\bibitem{deltamat} S. Fritsch, N. Kaiser and W. Weise, {\it Nucl. Phys.} {\bf
A750}, 259 (2005).\vs
\bibitem{2pidel} N. Kaiser, S. Gerstend\"orfer and W. Weise, {\it Nucl. Phys.} 
{\bf A637}, 395 (1998).\vs
\bibitem{spinstab} N. Kaiser, {\it Phys. Rev.} {\bf C70}, 054001 (2004).\vs
\bibitem{cola} G. Colangelo, J. Gasser and H. Leutwyler, {\it Nucl. Phys.} 
{\bf B603}, 125  (2001).\vs  
\bibitem{musch} M. Procura, B.U. Musch, T.R. Hemmert, and W. Weise, {\it Phys. 
Rev.} {\bf D75}, 014503 (2007).\vs
\bibitem{massi} M. Procura, B.U. Musch, T. Wollenweber, T.R. Hemmert, and W. 
Weise, {\it Phys. Rev.} {\bf D73}, 114510 (2006).\vs
\bibitem{hatsuda} N. Ishii, S. Aoki, and T. Hatsuda, {\it Phys. Rev. Lett.}
{\bf 99}, 022001 (2007); hep-lat/0710.4422.\vs 
\bibitem{dario} D. Vretenar, T. Niksic, and P. Ring, {\it Phys. Rev.} {\bf 
C68}, 024310 (2003).\vs 
\bibitem{fuchs} O. Plohl and C. Fuchs, {\it Nucl. Phys.} {\bf A}, (2007) in
print; nucl-th/0710.3700.\vs 
\bibitem{bentz} W. Bentz and A.W. Thomas, {\it Nucl. Phys.} {\bf A696}, 138 
(2001).\vs
\bibitem{finelli} P. Finelli, N. Kaiser, D. Vretenar and W. Weise, {\it Nucl. 
Phys.} {\bf A770}, 1 (2006).\vs
\end{thebibliography}
\end{document}